\documentstyle[11pt,aaspp4,flushrt]{article}

\newcommand{\cloudy}{\sc CLOUDY \rm}
\newcommand{\HST}{\it HST \rm}
\newcommand{\Strom}{Str\"{o}mgren}
\newcommand{\h}{$h^{-1}$}
\def\H0{$H_0 = 100h$ km s$^{-1}$ Mpc$^{-1}$}
\def\q0{$q_0 = 0.5$}
\def\O{{\rm O}}
\def\ion#1#2{{\rm #1\,\,\sc #2\rm}}
\newcommand{\OII}{[\ion{O}{ii}]}
\newcommand{\OIII}{[\ion{O}{iii}]}
\newcommand{\MgII}{\ion{Mg}{ii}}
\newcommand{\NeIII}{[\ion{Ne}{iii}]}
\newcommand{\NeV}{[\ion{Ne}{v}]}
\newcommand{\NII}{[\ion{N}{ii}]}
\newcommand{\Ha}{\mbox{{\rm H}$\alpha$}}
\newcommand{\Hb}{\mbox{{\rm H}$\beta$}}
\newcommand{\Hg}{\mbox{{\rm H}$\gamma$}}

\newcommand{\lamOII}{~$\lambda$3727}
\newcommand{\lamOIII}{~$\lambda$5007}
\newcommand{\lamMgII}{~$\lambda$2798}
\newcommand{\lamNeIII}{~$\lambda$3869}
\newcommand{\lamNeV}{~$\lambda$3426}
\newcommand{\lamNII}{~$\lambda$6584}
\newcommand{\pomega}{\mbox{$\varpi$}}
\newcommand{\A}{\mbox{$\Xi$}}

\newcommand{\W}{\mbox{$W_\lambda$}}

\begin{document}

\title{Emission Lines from Companion Galaxies of QSOs}

\author{Oleg Y. Gnedin}

\affil{Princeton University Observatory, Princeton, NJ~08544; \\
       ognedin@astro.princeton.edu}

\begin{abstract}
Anticipating new spectroscopic studies of quasar companions,
we calculate expected emission line spectra for the companion galaxies.
Under the assumption of isotropic photoionizing
radiation, we predict equivalent widths for \Ha, \Hb, \OIII \lamOIII,
\OII \lamOII, \NeIII \lamNeIII, and \MgII \lamMgII\ lines.
We consider 6 quasars, two of which (PKS 2135-14 and PKS 2349-014)
have very close companions with projected separation $< 4 h^{-1}$ kpc.
For these companions the calculated emission lines should be easily
observable, with $\W \gtrsim 20\ \AA$ for the \OIII, \OII, and \Ha\ lines.
Presence of strong emission lines from the companions occupying only a tiny
solid angle would suggest that 1) quasar radiation is isotropic;
2) companion galaxies have measurable amount of gas, which can be assessed
using the photoionization models; 3) host galaxies, if present, have been
depleted of gas or do not interact with the companions.
\end{abstract}


\section{Introduction}
There is increasing observational evidence that quasar activity
is associated with compact groups of galaxies (\cite{BC:91};
\cite{YE:93}, and references therein). Recent \HST observations
(\cite{BKS:95a},b, 1996; \cite{BKSS:97}, hereafter BKSS) 
of nearby luminous ($z<0.3$; $M_V \leq -22.9$)
quasars revealed extremely close companion galaxies, within
2\arcsec\ from the QSOs ($\sim$ 4\h\ kpc in projection).
Table \ref{tab:data} summarizes the known parameters for the companions
of six nearby quasars. The first three columns give quasar's name, its
redshift and absolute magnitude. The other columns describe the quasar
companions, including their redshifts, magnitudes, and
projected separations from the quasar, $d$,
both in arcseconds and kiloparsecs. The last column is the covering factor
$\pomega$, discussed in \S\ref{sec:lines}.

The physical conditions in the interstellar medium of the companions are
different from the Galaxy. Powerful ultraviolet radiation from the QSO,
if isotropic, is able to ionize a large fraction of the gas.
For a distant observer, the companion galaxies would look like giant HII
regions. Therefore, spectroscopic observations of the companion galaxies
of quasars directly test anisotropy of the quasar ionizing radiation
suggested by the implied AGN model (cf. for example \cite{A:93}).
The results of such observations may be important in understanding
the quasar environment.

\cite{SM:87} observed extended line emission around 
a sample of quasars in their narrow-band \OIII \lamOIII\ survey. 
In many cases, the emission regions are not confined to the nearest detected
companion and instead show asymmetric structures, such as bridges and tidal
tails. The Stockton-MacKenty results suggest that at least in some cases
companion galaxies are experiencing violent interactions
with the QSOs and that their gas may have been removed by interstellar shocks.
However, very little observational evidence exists on the emission-line spectra
of the close companions. Our work is intended to motivate
future observations by showing that gas content and metallicity can be
assessed by obtaining spectra of the companions.
We calculate the emission line spectrum expected from a single gas cloud
with various galactic parameters, which is illuminated by the hard UV radiation
from a central QSO source. Comparing our models with the 
results of narrow band imaging or high-resolution spectroscopy
yields estimates of the gas density and its distribution.

We outline in \S\ref{sec:lines} the calculations and the method that
can be used to study the ISM in the companion galaxies.
Next, we discuss in \S\ref{sec:results} the results for individual
companions. Finally, in \S\ref{sec:discussion},
we speculate on what can be learned about quasars and their environment
from the study of emission lines of the companion galaxies.

\section{Computing the emission lines}
\label{sec:lines} 
We use the photoionization program \cloudy, version 84.12 (\cite{cloudy})
to calculate the emission line spectrum of a cloud of gas exposed under
the quasar radiation. The density of the gas, $n_H$, is assumed constant
throughout the cloud. The shape of the incident QSO's spectrum is from
\cite{MF:87} and is normalized to the total quasar luminosity, $L_{QSO}$.
The underlying continuum of the companion is assumed to be the standard
spectrum of a Sbc galaxy, as reported by \cite{CWW:80}, normalized to
$L_{\rm comp}$. We explore five orders of magnitude in the gas density
(values characteristic of a normal galaxy), and four orders of magnitude in 
the ionizing flux. The equivalent widths of the emission lines are evaluated
relative to the outgoing continuum of the companion galaxy that includes
the component of the incident radiation absorbed and re-radiated by the gas.
For a similar exploration of clouds within, not outside, the AGN see
\cite{Betal:95}.

The gas in the cloud will be partially (or fully) ionized by the quasar
radiation and will resemble a giant HII region. The principle differences from
galactic HII regions are the geometry of the cloud and the shape of
the ionizing continuum. The companion
galaxy usually occupies a small solid angle as seen from the
quasar, and therefore only a fraction of the ionizing flux will hit
the inner surface of the cloud. This fraction is characterized by
the {\it covering factor}
\begin{equation}
\pomega \equiv {\Omega \over 4\pi} = {S \over 4\pi \, d^2},
\end{equation}
where $\Omega$ is the solid angle subtended by the cloud, $S$ is the
surface of the cloud closest to the quasar, and
$d$ is the distance to the quasar.

Unlike a galactic HII region, the ionization region around the quasar
does not have a definite boundary because of the power-law spectrum
of the ionizing radiation (\cite{BBK:71}; \cite{HS:83}).
The electron density declines smoothly
with radius and it is not possible to separate completely the ionized
and neutral zones. Nevertheless, it is instructive
to introduce a parameter analogous to the \Strom\ radius of the HII region.
Since \cloudy imposes spherical geometry on the cloud, the ionized region
is a spherical shell with surface area $S$ and mean radial width
$\Delta R_S$.
\begin{equation}
\Delta R_S = {Q({\rm H}) \, \pomega \over \alpha_B n_e n_p \, S \, f} \approx
2.7\; {Q_{55} \over f} \,
\left( {5\ {\rm kpc} \over d} \right)^2 \, \left( {1\ {\rm cm}^{-3} \over n_H}
\right)^2 \; {\rm kpc},
\label{eq:dRS}
\end{equation}
where $\alpha_B \approx 4 \times 10^{-13}$ cm$^3$ s$^{-1}$ is the
hydrogen recombination coefficient, and $n_e$ and $n_p$ are the electron and
proton densities, respectively. A filling factor, $f$, accounts for
the non-uniform distribution of the gas in the cloud.
$Q({\rm H})$ is the total rate of ionizing photons from the quasar.
For the chosen shape of the quasar continuum,
$Q_{55} \equiv Q({\rm H})/10^{55}\ {\rm s}^{-1} = L_{QSO}/3.2\times 10^{11}
 L_{\sun}.$ The ``\Strom\ width'', $\Delta R_S$, provides a useful
estimate of the effective volume occupied by the ionized gas that
determines the line emission strength from the cloud.

Table \ref{tab:ion} gives the values of $\Delta R_S$ for our
sample. Only two close companions have ionization zones comparable to their
size, and in the rest of the galaxies the \Strom\ zone is ionization limited
(its size $\Delta R > \Delta R_S$). The actual widths of the companions
(column 4), as well as their covering factors \pomega, were estimated
from the \HST images.

In the \cloudy calculations we use solar abundances of elements. Our choice
is supported by the recent result that the abundances of the heavy elements
considered here (O, N, Ne, and Mg) differ by less than 50\% in the solar
neighborhood (\cite{Eetal:93}), and by the current models of chemical
evolution of galaxies (for example, \cite{P:89}) that predict slow
evolution of metallicity over the last $5-10$ Gyr. External galaxies, however,
show a spread of metallicities. While including abundances as free parameters
would significantly complicate our models, it is still necessary to investigate
the extent to which varying metallicity of the gas may change the predicted
spectrum. Interstellar dust can also modify the thermal balance in the clouds
(\cite{SK:95}). We have conducted test calculations including the depletion
of certain elements onto dust grains (as described in \cite{Betal:91})
and varying metallicity parameter in a wide range of
$0.1 Z_{\sun} < Z < 5 Z_{\sun}$.

At very low metallicity ($Z \sim 0.1 Z_{\sun}$), the forbidden line
emission (for example, \OIII) increases as $Z$ rises, while the hydrogen
Balmer line intensity is essentially constant. Increasing metal abundance
leads to more efficient cooling and correspondingly lower electron
temperature. At about solar metallicity the temperature effect starts to
dominate and both \OIII\ and \Hb\ decline even as $Z$ increases
(\cite{Metal:85}). Including dust grains in the calculations gives rise to
the opposite phenomenon. The main effect of the grains is
depletion of metals from the interstellar medium (\cite{SK:95}). This
reduces cooling efficiency of the gas and leads to higher temperature
than in the dust-free case. Therefore the variation of the predicted
emission lines with $Z$ is reduced.

The test calculations have been performed for all quasar companions considered
in \S\ref{sec:results}. In the range $0.3 < Z/Z_{\sun} < 3$, the intensity
of the \OIII\ line varied by an average of 45\% for the two close companions
and by 53\% for the other companions with lower ionization level.
In the range $0.1 < Z/Z_{\sun} < 5$ for the close companions, \OIII\ was
reduced by about 90\% compared to its value at
$Z \approx Z_{\sun}$. The \OIII/\Hb\ line ratio changes more slowly
than the \OIII\ intensity at high metallicity because of similar drop
in \Hb. In all calculations involving the two close companions the average
electron temperature was in the range $T_e = 8\times 10^3 - 3\times 10^4$ K,
and for the low-ionization companions $T_e = 6\times 10^3 - 2\times 10^4$ K.
From these tests we conclude that choosing solar abundances maximizes
the forbidden line emission from the quasar companions.

\subsection{Distribution and density of gas in the close companions}
\label{sec:method_close}
When companions are very compact or very close to the quasar, such that
$\Delta R < \Delta R_S$, essentially all gas is ionized and the intensity
of the emission lines is directly proportional to the effective
volume occupied by the gas, $V = S\times \Delta R\times f$.
The unknown parameters are the gas density $n_H$ and the filling factor $f$.
We have computed a two-dimensional set of ionization models, varying $n_H$
from $10^{-2}$ to $10^3$ cm$^{-3}$, and $f$ from 1 to $10^{-4}$.

Figure \ref{fig:map_m} shows the contours of the equivalent widths
for the \Ha, \Hb, \OIII \lamOIII, \OII \lamOII, \NeIII \lamNeIII, and
\MgII \lamMgII\ lines, corresponding to the rest frame of the cloud.
The \OIII\ line illustrates the transition from the matter-bounded to
ionization-bounded models. On the left side of the plot
$\Delta R_S > \Delta R$ and the contours of \W\ coincide with the lines
of constant \Strom\ radius ($f \propto n_H^{-2}$). From the simple
three-level model of an oxygen atom we can predict $\W(\OIII) \propto n_H^3 V$.
For large densities, $\Delta R_S$ drops below $\Delta R$ and the line
intensity becomes essentially independent of the amount of gas, $f$.
Also as density increases, the ionization level of oxygen decreases
and the \OII\ line becomes stronger.

In the \Ha, \Hb, and \NeIII\ panels there is a plateau on the high-density
side where the equivalent widths are essentially constant. Here all of
the ultraviolet photons are trapped inside the cloud, and the ionization
zone is limited by the quasar intensity. For example, \W(\Hb) cannot
rise much more than 30 \AA\ for any combination of the gas density
and the filling factor.

Note that \OIII\ $\lambda 4959$ line is also present if \OIII \lamOIII\
is observed. We do not include this line
because of the approximate relation $\W(5007) \approx 3\times \W(4959)$
for $T \sim 10^4$ K (\cite{O:89}).
If the two lines are not resolved, the equivalent
width of the \OIII \lamOIII\ line should be corrected accordingly.

The intersection of the isocontours of all the observed \W\ gives the 
point on the $f - n_H$ diagram that corresponds to the ionization model best
reproducing the observations. Besides an estimate of $n_H$, we obtain also
the full emission-line spectrum of the object (within the simple,
constant density model). Therefore we predict the intensity and equivalent
widths for other strong lines, and these in turn can be tested by future
observations.
Given good data from several lines (including upper limits),
it seems possible to construct a model with a realistic gas distribution
covering a range of densities $n_H$.

\subsection{Density of gas in the ionization-bounded companions}
\label{sec:method_far}

If the companions are sufficiently far from the quasar, the strength
of the emission lines is limited by the number of ionizations. At high
enough density, the effective volume of the ionized region is
$V = S\times \Delta R_S\times f$ and the equivalent widths of the hydrogen
lines are proportional to a single ionization parameter
\begin{equation}
\A \equiv {L_{QSO} \, \pomega \over L_{\rm comp}}.
\label{eq:A}
\end{equation}
This parameter is, in principle, directly observable.
Values of \A\ for the companion galaxies in
our sample are given in the last column of Table \ref{tab:ion}.
It could be related to the usual ionization parameter $U$:
$\A = U \times n_H({\rm cm}^{-3}) S({\rm kpc}^2) (L_{\rm comp}/8.8\times 10^9\
 L_{\sun})^{-1}$.
The two parameters are similar in value but \A\ has the advantage of being
independent of the gas density. This allows us to separate the intensity
of the incident flux from the parameters intrinsic to the companion.

Figure \ref{fig:map_i} shows the results of our calculations for the 
same emission lines, now as function of \A\ and $n_H$.
The two hydrogen lines, \Ha\ and \Hb, demonstrate the simple variation
of the equivalent widths. When the density is large,
$\W(\Ha,\Hb) \propto \A$ almost independently of $n_H$.
On the other hand, for $n_H < 1$ cm$^{-3}$ the ionization is complete
and $\W(\Ha,\Hb) \propto n_H^2$.
The corners of the contours lie on a single line $\Delta R = \Delta R_S$
(i.e. $\A \propto n_H^2$). Note that both permitted and forbidden lines
show a general break along that line.

Behavior of the oxygen lines can be understood using the three-level
model of the oxygen atom. The dominant excitation mechanism is collisions
with electrons unless the radiation field is so strong that recombination
to excited levels becomes important. For the \OIII\ line, collisional
excitation prevails if $n(\O^{+2})/n(\O^{+3}) > 0.0025$, which is
always satisfied in our models. We have also checked that the quasar
radiation is strong enough that photoionization of neutral oxygen
atoms wins over the charge exchange with protons (see \cite{Sp:78})
if $n_p \lesssim 76\ Q_{55}\ (d/5\ {\rm kpc})^{-2}$ cm$^{-3}$. 

For low densities ionization is so high that the ion abundances of
the heavy elements are controlled by recombination from the next
ionization state. As a result, all equivalent widths follow the
approximate relation $\W({\rm O, Ne, Mg}) \propto n_H^3 \A^{-1}$.
In the limit of a weak radiation field (or high density), the
ion concentrations
scale differently and so do the equivalent widths of the lines.
The ionization zone is limited by $\Delta R_S$, and
$\W(\OII) \propto \A^2 n_H^{-1}$, $\W(\OIII) \propto \A^3 n_H^{-2}$.

Figure \ref{fig:map_i} allows us to find the best ionization model
describing a set of the observed emission lines.
The ionization parameter of the cloud could be determined directly
from the observations (see eq. [\ref{eq:A}]).
Ideally the contours of the equivalent widths
should all intersect in one point with the horizontal line of the
known \A, and even one measured line would suffice
to obtain the value of the gas density.
However given the uncertainty of measurements, it is desirable to have
several lines in order to better constrain the estimate of $n_H$.

How does the observational uncertainty in \W\ propagate to the
inferred values of $n_H$? We have considered a range of values of
\W\ and obtained the corresponding range of allowed densities. Our
tests show that the relative error in $\log n_H$ is of the same order
as that in $\log{\W}(\OIII,\OII)$, with
$\Delta \log n_H \approx 1.3 \Delta \log{\W}$ for a wide region on
the $\A - n_H$ diagram. Therefore, observational errors
are not strongly enhanced in the estimate of the gas density.

\section{Discussion of individual galaxy companions}
\label{sec:results}

In this section, we apply the results of our calculations to the individual
companions listed in Table \ref{tab:data}. 
In calculation of luminosities of quasars and their companions we use
a solar V magnitude of $M_{V,\sun} = 4.84$ (\cite{Wo:94}) and
the Hubble constant \H0.
Note that choice of $H_0$ does not affect the predicted equivalent widths
because the ionizing fluxes from the quasars are invariant to distance scaling.

To account for the third spatial dimension, we correct the projected
separations of the companions from the quasars by a factor of
$\sqrt{3/2} \approx 1.22$. These larger values are used in calculation
of the ionization models. The second column of Table \ref{tab:ion}
gives the corrected separations for all of the companions.

\subsection{Close companions}
\label{sec:close}

\subsubsection{PKS 2135-14}
\label{sec:pks2135}
\cite{S:82} presented a spectrum and a detailed discussion of the
closest companion to PKS 2135-14.  Its compact nature is best revealed
on the \HST image (BKSS), where the companion is unambiguously
distinguished from the nebulosity surrounding the quasar.  Based
on the spectrum of the companion, \cite{CS:97} showed that it is
actually a foreground star.  Therefore, the narrow line emission
is produced by the gas in the fuzzy nebulosity extending over more
than $8h^{-1}$ kpc.  Our ionization models describe the state of
this extended gas.

We read the equivalent widths of the \OIII, \OII, and \Hb\ lines
by visual inspection of the spectrum in \cite{S:82}.
Table \ref{tab:closecomp} shows our estimates of $\W$;
uncertainty in the values should be taken to be at least 20\%.
We construct the best fitting photoionization model using the calculations
described in \S\ref{sec:method_close}.
Figure \ref{fig:pks2135} shows contours of \W\ for the three measured lines.
One striking feature of the plot is that
all three contours run reasonably close along the line corresponding to the
constant \Strom\ radius ($f \propto n_H^2$). Therefore
any model along the line is, in principle, capable of reproducing
the observed emission. However, the isocontours never cross
in one point, suggesting that more than one constant density model
is required for the adequate fit of observations, provided
the estimated equivalent widths are correct.

We considered a simple combination of high and low density clouds
(see Table \ref{tab:closecomp}).  The high density clouds are needed
to reproduce the \OII\ emission, since most of the oxygen is doubly
ionized at small densities ($n_H < 3$ cm$^{-3}$) at the apparently
small observed separation from the quasar.  The low density clouds
are likely to be present in the large extended nebulosity.  We stress
that the choice of high and low density clouds is not uniquely set by
the available data.

The two models together produce the right amount of the \OII\ emission,
23\% less of the \OIII, and 50\% more of the \Hb\ than is estimated from
the observed spectrum. The \Ha\ line should be conspicuous, with
$\W \sim 24\ \AA$, but it lies outside the spectral range of Stockton's data.
The \NeV\ \lamNeV\ line becomes moderately strong if we take any lower density
for the model $a$. However it is barely detectable in the spectrum,
and the estimated upper limit on its equivalent width
is about 3 \AA. Thus the gas density of the model $a$ is limited to
$n_H \gtrsim 2.5$ cm$^{-3}$. On the other hand, a larger filling factor
($f > 10^{-2}$) leads to a significant overestimate of the \Hb\ line while
still not producing enough of the \OIII.

Another companion is located 6\arcsec\ to the East from PKS 2135-14.
\cite{CS:97} confirmed that it is a galaxy at the redshift of the quasar,
however its spectrum does not show strong emission lines.

\subsubsection{PKS 2349-014}
\label{sec:pks2349}
The closest companion of PKS 2349-014 exhibits a tidal interaction
with the host QSO (see Bahcall et al. 1995a).  Its redshift has been
confirmed by \cite{Metal:96}.  The estimated three-dimensional distance
from the quasar is 4.2\h\ kpc.  There are also apparent tidal tails and
a large nebulosity surrounding, but offset from, the quasar.  If the
quasar has a host galaxy similar to the Milky Way, the companion would
inspiral due to dynamical friction in $7\times 10^8$ yr.

We have computed two illustrative models with plausible gas densities
of $n_H = 1$ cm$^{-3}$ and 4 cm$^{-3}$, respectively.
We expect that the highly ionized, and therefore heated,
gas would tend to be evenly distributed over the volume of the companion.
Because of the faint continuum of the companion,
the predicted equivalent widths in the second model are very large
(see Table \ref{tab:closecomp}). For example, $\W(\OIII) \approx 250\ \AA$.
Spectroscopic observations of the companion galaxy are necessary to
constrain the ionization model and to estimate the present amount of gas.

Because of the extreme proximity of the companion, radiation pressure
from the quasar may affect the distribution of the gas.  We use \cloudy\
calculations to estimate the radiative acceleration on a hydrogen atom
in the companion.  We assume an isothermal distribution of mass of the
companion and adopt the mass-to-light ratio of $M/L_V=1$ characteristic
of moderately young stellar population.  We find that for the models
I and II, gravity of the companion dominates the radiation pressure
out to 18 kpc and 8 kpc, respectively, much farther than the actual
extent of this object.  Therefore, the gas clouds are not driven out
of the companion.  We have checked that the momentum transfer due to
absorption of ionizing radiation by hydrogen atoms (for example,
\cite{H:95}) is at least an order of magnitude weaker than the force
due to resonant Ly$\alpha$ scattering.

\subsection{Ionization-bounded companions}
\subsubsection{NAB 0205+02}
\label{sec:nab0205}
NAB 0205+02 from the survey by \cite{BBS:73} has a mysterious companion
at a projected separation of 22\h\ kpc;
this companion is very compact but is bright in the \OIII\ line.
\cite{SM:87} estimated the equivalent width of \OIII\ line of at least
400 $\AA$ from the narrow line image. On the other hand, the object
was not detected at all on their line-free continuum image.
Recent \HST\ F606W broad band observation of the companion (BKSS),
which includes the \OIII\ line, shows that the image consists of two
tiny clumps
connected to each other by a bridge. There is no evidence on the \HST
image that the quasar host extends all the way to the companion,
suggesting that the gas in the object is not affected by direct
interaction with the host.
The line emission in the companion may, in principle, be caused by
photoionization or collisional excitation by sources internal to
the companion. We assume here that it is not the case and discuss instead
the situation when photoionization by the quasar is the dominant source
of energy for the emission lines.

The \Strom\ radius for the companion (see Table \ref{tab:ion}) 
is less than the apparent extent of the object
($\sim$ 3.5 kpc in NS direction, and 0.5 kpc in SW direction),
so that the gas will be only partially ionized
for $n_H > 0.3$ cm$^{-3}$. The amount of \OIII\ emission is therefore
determined by the ionization parameter \A, which
in turn depends on the covering factor. Assuming that the size of the object
along the line-of-sight is the same as in the NS direction, we get an upper
limit on \pomega\ of $1.4\times 10^{-3}$, or $\A \approx 1$.
Using the estimated three-dimensional separation from
the quasar of 26.5 kpc, we found that no simple ionization model,
with the ionizing source centered on the quasar, yields
an equivalent width of the \OIII\ line as large as is observed.
The smallest covering factor needed to produce $\W(\OIII) \gtrsim 400 \AA$
is $\pomega = 2.4\times 10^{-3}$, and the gas density in this case is
0.3 cm$^{-3}$.

Can the emitting gas be distributed more broadly than the \HST image shows?
Apparently, the narrow-line image of Stockton \& MacKenty (1987;
their Figure 2e) is large
but this could be the effect of lower resolution relative to \HST.
Nonetheless, we present
the ionization model with $\pomega = 2.4\times 10^{-3}$
that best fits the observations. If the model is confirmed by new spectroscopic
data, a better understanding of the distribution of gas
and stars in the companion will be required.
If not, a much more elaborate ionization model
will be needed to describe this unusual object.

The one-cloud ionization model predicts large equivalent widths for
virtually all of the emission lines listed in Table \ref{tab:farcomp}.
The \OII, \Ha, \Hb, \NeIII, and \MgII\ lines have calculated \W\
in excess of 40 \AA. Also included in Table \ref{tab:farcomp} are strong
\NII\ and \Hg\ lines.
Our simplest model gives an upper limit
(for $f=1$) on the total mass of the gas in the companion of
$M_{gas} \lesssim 8\times 10^7\, M_{\sun}$.

The faint continuum of the companion suggests the possibility that most of
the light might be produced by thermal bremsstrahlung rather than starlight.
The bremsstrahlung luminosity of the companion would be (\cite{O:89})
\begin{equation}
L_{ff} = 1.42\times 10^{-27} g_{ff} T^{1/2} n_e n_i Z^2 V\, {\rm erg\ s}^{-1}
  \approx 1.5\times 10^7 L_{\sun} \, T_4^{1/2} n_e n_i Z^2,
\end{equation}
where $g_{ff} \sim 1.3$ is the free-free Gaunt factor, $n_i$ is the density
of ions of charge $Z$, and $V \sim 10$ kpc$^3$ (for 
$\pomega = 2.4\times 10^{-3}$) is the volume of the gas.
The observed companion luminosity is $L_{\rm comp}=1.3\times 10^8
L_{\sun}$ (BKSS), and therefore one needs a density of $n_e > 3$ cm$^{-3}$
to produce the right amount of emission.
This in turn requires a larger covering factor of
$\pomega > 6\times 10^{-3}$
in order to match the observed \W(\OIII) with the gas density larger
than 3 cm$^{-3}$. As noted before, the covering factor can hardly
be larger than $(2-3) \times 10^{-3}$.
Also, self-gravity of the gas will not be strong enough to resist
the radiation pressure force, as indicated by the \cloudy calculations.
Therefore the continuum of the companion cannot be accounted for by
thermal bremsstrahlung only. Most likely, the detected emission
arises from a combination of emission lines and faint starlight.

\subsubsection{PHL 909}
Galaxy F in the \cite{BKS:96} list of objects around PHL 909
is 65\h\ kpc away from the quasar.
\cite{B:96} has estimated the equivalent widths of the \OII\ and \OIII\
lines as 10 \AA\ and 2 \AA, respectively. 

Figure \ref{fig:phl909} shows the corresponding contours of the lines
as function of \A\ and $n_H$, using the calculations from
\S\ref{sec:method_far}. The intersection of the two lines gives a slightly
larger covering factor than is estimated from the \HST\ image, so that
we had to lower the gas density in order to produce the observed \W.
The best ionization model is obtained with $n_H = 0.8$ cm$^{-3}$ for
$\pomega = 5\times 10^{-3}$. Because of the large distance to the quasar,
the ionization flux is not very powerful and there are no strong lines
predicted. Besides the oxygen lines, there could be detectable
\Ha\ and \NII \lamNII\ lines both with $\W \sim 4\ \AA$.

\subsubsection{HE 1029-1401}
\cite{W:94} showed that the quasar HE 1029-1401 is surrounded
by a loose group of galaxies with strong emission lines.
Table \ref{tab:farcomp} gives the equivalent widths for
the \OIII, \OII, and
\Hb\ lines for the closest galaxy with a redshift similar to the quasar.
The companion is at deprojected distance of about 150\h\ kpc, and the
covering factor is small ($\pomega \sim 6\times 10^{-4}$). Even with
$\pomega = 10^{-3}$ and $n_H = 1$ cm$^{-3}$, photoionization
by the quasar is too weak by an order of magnitude to produce the measured
lines (Table \ref{tab:farcomp}).

The observed emission can be explained by star formation within
the companion galaxy. The continuum level at \Hb\
($\lambda = 4861 \AA$) is related to the total luminosity $L_{\rm comp}$
using the standard
spectrum of an Sbc galaxy from Coleman et al. (1980); we find
$\lambda L_\lambda(\Hb) = 0.93\, L_{\rm comp}$. The observed equivalent
width of the \Hb\ line gives then the intensity of the
line, $L(\Hb) = 2.1\times 10^{40} h^{-2}$ erg s$^{-1}$.
Using \cite{K:83} relation for the star formation rate (SFR) of spiral
galaxies to the luminosity of the \Ha\ line and
assuming $L(\Ha) \approx 3\times L(\Hb)$,
we find SFR $\approx 0.6\ M_{\sun}\, {\rm yr}^{-1}$, which can be achieved
by a normal size spiral galaxy.

\subsubsection{PKS 0405-123}
The companion galaxy to PKS 0405-123 has been observed by \cite{MS:85},
and its redshift has been confirmed by \cite{EY:94}.
The spectrum of the galaxy has a strong
\OII\ emission line, with the estimated equivalent width of 
$\W(\OII) \sim 12\ \AA$. 
The luminosity of the companion ($M_V = -20.2$) suggests that it may be
a large spiral galaxy. There is no \HST observation of this
quasar field and the available image from \cite{MS:85} does not permit
robust determination of the covering factor of the companion. 
We estimate the size of the galaxy to be 10 kpc, and assume it is almost
face-on towards the quasar. This gives $\pomega \sim 3\times 10^{-3}$.
Using Figure \ref{fig:map_i}, we estimate the gas density to be about 
8 cm$^{-3}$. The ionization model predicts some measurable emission
in \Ha, \MgII\ and \NII, with $\W \gtrsim 5\ \AA$.
If the companion is inclined with respect to the quasar,
\pomega\ will be smaller and the ionizing flux from the quasar will not be
strong enough to produce the observed \OII\ emission.

\section{Summary}
\label{sec:discussion}
The primary goal of this paper is to encourage new spectroscopic
observations of quasar companions in order to learn more about the
companion galaxies, isotropy of the quasar emission, and the relation
between the galaxies and the quasars. We have calculated the equivalent
widths of the strongest emission lines expected to be produced by quasar
photoionization of close galactic companions and applied our models to
the six companions with the observed lines.

\begin{figure}
\plotfiddle{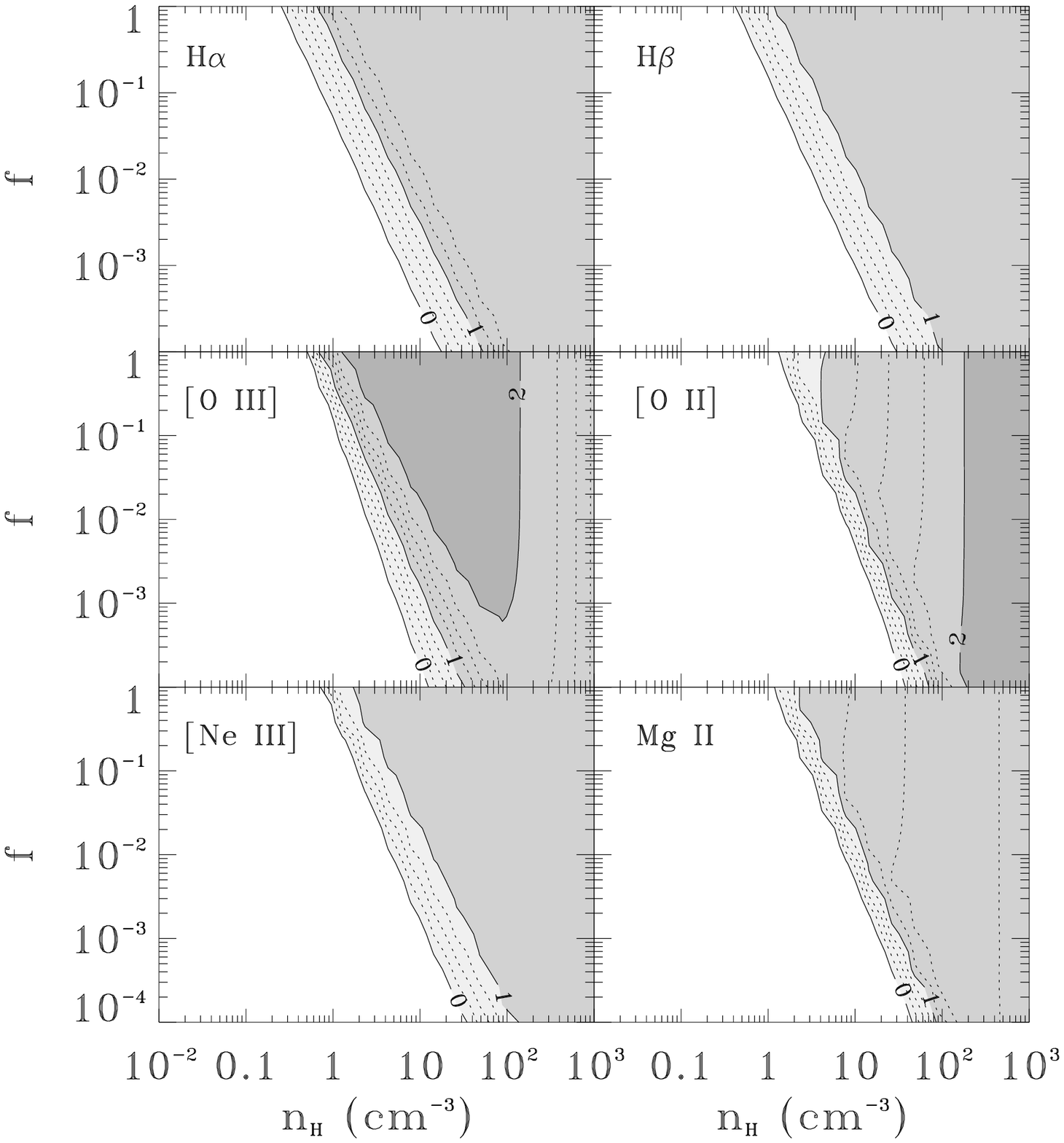}{6.9in}{0}{90}{90}{-270}{-108}
\caption{Contour plot of equivalent widths for the \Ha, \Hb, \OIII \lamOIII,
\OII \lamOII, \NeIII \lamNeIII, and \MgII \lamMgII\ emission lines
as function of the filling factor, $f$, and density, $n_H$,
of gas in a close companion. Solid lines are the contours of $\W=1$, 10,
100 \AA, with the numbers labeling $\log{\W}$. Dotted lines represent 0.25 dex
steps. The other fixed parameters of the calculation are:
$L_{QSO}=2\times 10^{11}\, L_{\sun}$, $L_{\rm comp}=5\times 10^9\, L_{\sun}$,
$d=3.2$ kpc, $\Delta R=2$ kpc, $\pomega=0.02$.\label{fig:map_m}}
\end{figure}

\begin{figure} 
\plotfiddle{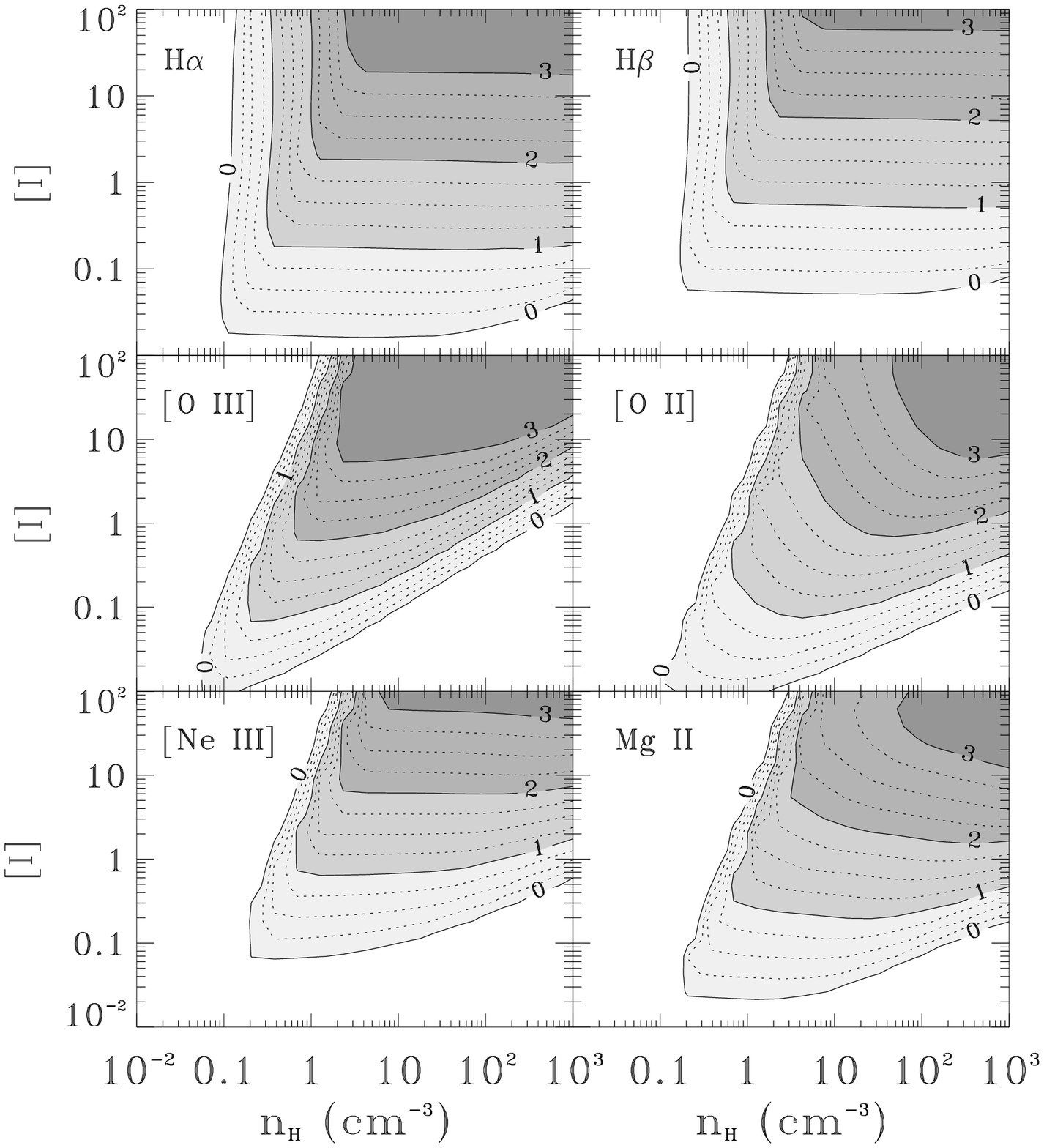}{6.9in}{0}{90}{90}{-270}{-108}
\caption{Contour plot of equivalent widths for the \Ha, \Hb, \OIII \lamOIII,
\OII \lamOII, \NeIII \lamNeIII, and \MgII \lamMgII\ emission lines
as function of the ionization parameter, \A, (eq. [\protect\ref{eq:A}])
and density of the gas, $n_H$, for the ionization-bounded quasar companions.
Solid lines are the contours of $\W=1$, 10, 100, and 1000 \AA, 
with the numbers labeling $\log{\W}$. The dots represent 0.25 dex
steps. The other fixed parameters of the calculation are:
$L_{QSO}=2\times 10^{11}\, L_{\sun}$, $L_{\rm comp}=2\times 10^8\, L_{\sun}$,
$\Delta R=1$ kpc, $f=1$.\label{fig:map_i}}
\end{figure}

\begin{figure}
\epsscale{0.53}
\plotone{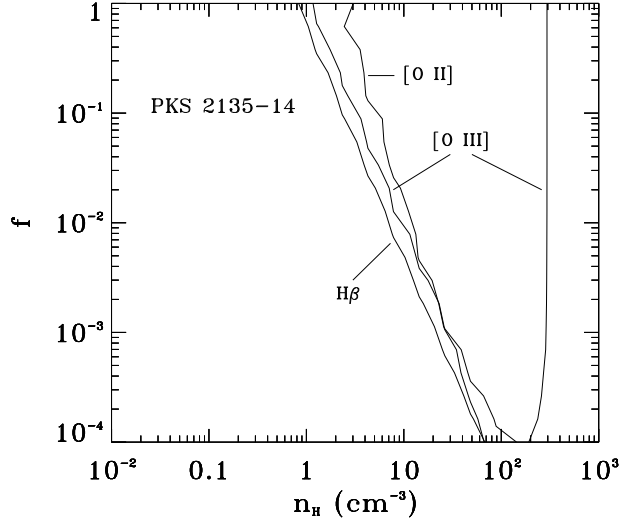}
\caption{Example of estimating the filling factor $f$
and density $n_H$ of gas around PKS 2135-14,
using the map of ionization models from Figure \protect\ref{fig:map_m}.
Solid curves are the isocontours of the observed equivalent
widths for the three emission lines: $\W(\OIII) = 69\ \AA$,
$\W(\OII) = 8\ \AA$, and $\W(\Hb) = 5\ \AA$.
Intersection of all the lines gives the best ionization model.
In this case, however, \OIII\ and \Hb\ contours are almost parallel to each
other with some offset from \OII.\label{fig:pks2135}}
\end{figure}

\begin{figure}
\epsscale{0.53}
\plotone{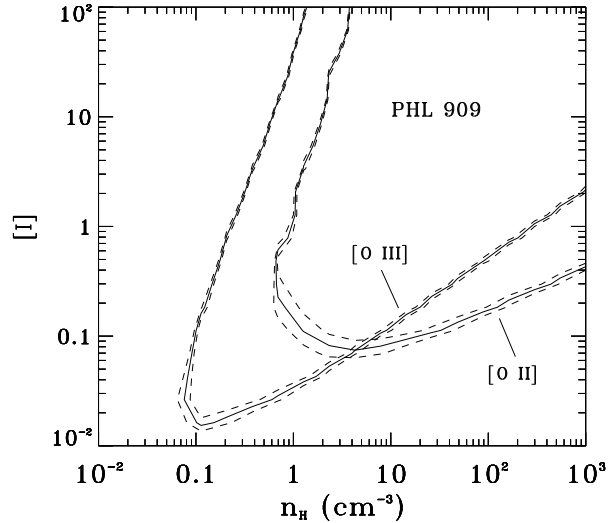}
\figcaption{Isocontours of the emission lines for the companion of PHL 909:
$\W(\OII) = 10\ \AA$ and $\W(\OIII) = 2\ \AA$. Intersection of the two lines
gives a higher density than in the best-fitting model ($n_H = 0.83$ cm$^{-3}$)
because of the smaller observed \A\ parameter.
The dashes outline the region in this two-dimensional space that is
allowed by uncertainty in \W\ of 30\%.
\label{fig:phl909}}
\end{figure}

In our exploratory calculations we have assumed constant density gas
clouds and solar abundancies of the elements. Choice of the solar
metallicity maximizes the forbidden line emission. For the close companions
of PKS 2135-14 and PKS 2349-014, the predicted equivalent widths
are very strong, $\W \gtrsim 20\ \AA$. Our results suggest
(see Figures \ref{fig:map_m} and \ref{fig:map_i}, and Tables
\ref{tab:closecomp} and \ref{tab:farcomp}) that a narrow
range of gas densities of $n_H$ around 1 cm$^{-3}$ may produce the observed
emission lines. In three cases (the two close companions mentioned above and
the exotic companion of NAB 0205+02) the models constrain the amount of gas
to be of order $M_{gas} \sim 10^7 - 10^8\ M_{\sun}$.

Determining the amount of gas in or around the close companions
is important for understanding the dynamical interactions of quasars
with their environment. \HST\ images (e.g.,
Bahcall et al. 1995a and BKSS)
show clearly that, at least in some cases, quasars are
interacting violently with their galactic companions.
The long tidal tail observed near PKS 2349-014 suggests the presence of a
massive host galaxy. The small
quasar-companion separation of $\sim 4$\h\ kpc indicates that the companion
is well inside the typical scale of a large galaxy. Gas shocks
in the interstellar medium of the host might be expected to remove all
of the gas from the small companion.
Since the radiation pressure of the quasar is not strong enough to push
away all the gas of the companion, absence
of the strong emission lines predicted by our isotropic photo-ionization
models could suggest an anisotropic ionizing flux or shock interactions
with the massive host.\footnotemark\
On the other hand, if strong photoionized emission lines are observed,
a possible explanation might be an almost ``naked'' central quasar,
which accretes gas from disrupted dwarf companions.
This possibility has been suggested recently by Bahcall et al. (1995a)
and by Fukugita \& Turner (1996).

\footnotetext{
An alternative explanation for the lack of line emission in the companions
(proposed by J. Goodman) could be a very short lifetime of the quasar. If the
central engine is powered by the disrupted material of a single star
and lasts for only several years, the ionizing radiation from the quasar
may have not yet reached the companions some 10,000 light years
away.}

The degree of anisotropy of the quasar ionizing radiation, which is important
for understanding the unified AGN model (e.g. \cite{A:93}), can be tested
directly. If the continuum emission from
the quasar is concentrated in narrow beams, the small solid angle
of the companion galaxies makes them unlikely to be ionized.
The companion of NAB 0205+02
presents a particularly strong test.
Its tiny covering factor ($\pomega \sim 10^{-3}$)
guarantees that the companion will not be ionized unless the quasar
radiation is isotropic. Spectroscopy of that object
can determine if the line emission is consistent with
photoionization by the quasar.

The gas in the host galaxy itself should be strongly ionized by the quasar
and all of the emission lines considered
in this paper are expected to be observable. For the simplest model
of a disk galaxy with the constant gas density, the effective radius
of the ionization zone is
$R_{ion} \approx 4.0 \, (L_{11}/f n_H^2)^{1/3}$ kpc, where
$L_{11} \equiv L_{QSO}/10^{11} L_{\sun}$.
As noted by \cite{B:85} and \cite{Cetal:87}, total depletion of gas
and dust occurs in the inner few kiloparsecs. Thermal expansion of the gas
in the direction perpendicular to the galactic plane leads to an
amorphous envelope of hot gas around the quasar extending up to
several tens of kpc. \cite{SM:87} found such large emission envelopes
to be ubiquitous around quasars. Therefore the period of quasar activity
should be longer than the timescale for thermal expansion and
subsequent cooling of the gas, or $\sim 10^8$ years.

Close companions to three other quasars, PKS 1302-102,
3C 323.1 (=Q1545+210) and PG 1202+281, have been detected with \HST
(\cite{BKS:95b})
but no spectra are currently available.
\cite{SM:87} found an extremely strong \OIII\ emission around Q1545+210,
but they could not unambiguously associate the emission with the companion
7 kpc away from the quasar.
We would like to attract reader's attention to the 6 companions discussed
in this paper and the other 3 objects mentioned above.
They present a new interesting opportunity to understand quasars.

\acknowledgments
This project would not be possible without support and encouragement
from Professor John Bahcall.
We gratefully acknowledge use of the Gary Ferland's program \cloudy.
Don Schneider provided a computer readable version of the spectrum
of Sbc galaxy from Coleman et al. (1980).  T. Boroson generously measured,
and allowed us to use here, the equivalent widths of the \OII\ and \OIII\
emission lines on his spectrum of galaxy F in the field of PHL 909.
We thank J. Bahcall, B. Draine, G. Ferland, J. Goodman, J. Ostriker,
M. Rees, and L. Spitzer for helpful discussions.
This project was supported in part by NSF grant AST-9424416.

\begin{deluxetable}{lccccccc}
\tablecaption{Photometric Parameters of Companion Galaxies\tablenotemark{a}
              \label{tab:data}}
\tablecolumns{8}
\tablehead{\multicolumn{3}{c}{Quasar} & \multicolumn{5}{c}{Companion} \nl
           \cline{1-3} \cline{4-8}
           & \colhead{$z$} & \colhead{$M_V$} & \colhead{$z$} & \colhead{$M_V$}
           & \colhead{$d$(\arcsec)} & \colhead{$d$(\h\ kpc)} & 
           \colhead{\pomega\tablenotemark{b}}}
\startdata
PKS 2135-14  & 0.200 & -23.5 & 0.200 (?) & -19.4 &  2.0 &   4.2
             & $\lesssim 2\times 10^{-2}$ \nl
PKS 2349-014 & 0.173 & -23.4 & 0.173     & -17.7 &  1.8 &   3.4
             & $7\times 10^{-3}$ \nl
NAB 0205+02  & 0.155 & -23.0 & 0.155     & -15.4 & 12.0 &  21.6
             & $\lesssim 1.4\times 10^{-3}$ \nl
PHL 909      & 0.171 & -22.9 & 0.169     & -20.1 & 28.2 &  53
             & $5\times 10^{-3}$ \nl
HE 1029-1401 & 0.086 & -23.2 & 0.086     & -18.5 & 112  & 120
             & $6\times 10^{-4}$ \nl
PKS 0405-123 & 0.574 & -24.3 & 0.570     & -20.2 & 13   &  49
             & $3\times 10^{-3}$
\enddata
\tablenotetext{a}{Using \H0, \q0.}
\tablenotetext{b}{Covering factor \pomega\ of the companions, estimated from 
  their \HST images.}
\end{deluxetable}

\begin{deluxetable}{lcccc}
\tablecaption{Ionization Parameters of Companion Galaxies\label{tab:ion}}
\tablecolumns{5}
\tablehead{& \colhead{$d'$ (\h\ kpc)\tablenotemark{a}} &
           \colhead{$\Delta R_S$ (kpc)\tablenotemark{b}} & 
           \colhead{$\Delta R$ (\h\ kpc)} & \colhead{\A}}
\startdata
PKS 2135-14  &  5.1 & 1.5                 &   3 & 0.9  \nl
PKS 2349-014 &  4.2 & 2.3                 &   1 & 1.3  \nl
NAB 0205+02  & 26.5 & $4.1\times 10^{-2}$ & 0.5 & 1    \nl
PHL 909      &   65 & $5.9\times 10^{-3}$ &   9 & 0.07 \nl
HE 1029-1401 &  147 & $1.6\times 10^{-3}$ &   6 & 0.05 \nl
PKS 0405-123 &   60 & $2.7\times 10^{-2}$ &  10 & 0.1
\enddata
\tablenotetext{a}{Separation from the quasar corrected for the projection
effect; $d' \equiv d \times \sqrt{3/2}$.}
\tablenotetext{b}{For $n_H = 1$ cm$^{-3}$, $f=1$. The \Strom\ radius
scales as $\Delta R_S \propto n_H^{-2} f^{-1}$ (see eq.[\ref{eq:dRS}]).}
\end{deluxetable}

\begin{deluxetable}{lcccccc}
\tablecaption{Emission lines for close companions\label{tab:closecomp}}
\tablecolumns{7}
\tablehead{& \multicolumn{4}{c}{PKS 2135-14} & \multicolumn{2}{c}{PKS 2349-014}
           \\ \cline{2-5} \cline{6-7}
           & & \colhead{Model $a$} & \colhead{Model $b$} \\
           \colhead{\W(\AA)} & \colhead{[1]} & \colhead{(low-density)} &
           \colhead{(high-density)} & \colhead{total} &
           \colhead{Model I} & \colhead{Model II}}
\startdata
\OIII \lamOIII    &       69  &   9.8  &   43.3  &  53  &    53  &  247  \nl
\Hb               &        5  &   2.0  &    5.6  &   8  &     8  &   24  \nl
\OII \lamOII      &        8  &  $<1$  &    7.9  &   8  &  $<1$  &   19  \nl
\Ha               &  \nodata  &   6.3  &   17.4  &  24  &    26  &   73  \nl
\MgII \lamMgII    &  \nodata  &  $<1$  &    3.4  &   3  &  $<1$  &   25  \nl
\NeIII \lamNeIII  &  \nodata  &  $<1$  &    3.8  &   4  &     4  &   22  \nl
\NeV \lamNeV      &  \nodata  &   2.3  &   $<1$  &   2  &    25  &    6  \nl
\cutinhead{Parameters of ionization models}
$n_H$ (cm$^{-3}$) &           &   3    &     45  &      &     1  &    4  \nl
$f$               &        & $10^{-2}$ & $10^{-4}$ &    &     1  &    1  \nl
$M_{gas}$ ($M_{\sun}$) & & $>5.4\times 10^6$ & $8.1\times 10^5$ & 
   $6.2\times 10^6$ & $2.4\times 10^7$ & $9.6\times 10^7$
\enddata
\tablerefs{[1] \cite{S:82}.}
\end{deluxetable}

\begin{deluxetable}{lcccccccc}
\tablecaption{Emission lines for ionization-bounded companions
              \label{tab:farcomp}}
\tablecolumns{9}
\tablehead{& \multicolumn{2}{c}{NAB 0205} & \multicolumn{2}{c}{PHL 909} &
           \multicolumn{2}{c}{HE 1029} & \multicolumn{2}{c}{PKS 0405} \\
           \cline{2-3} \cline{4-5} \cline{6-7} \cline{8-9}
           \colhead{\W(\AA)} & \colhead{[1]} & \colhead{Model} & 
           \colhead{[2]} & \colhead{Model} & \colhead{[3]} & 
           \colhead{Model} & \colhead{[4]} & \colhead{Model}}
\startdata
\OIII \lamOIII   & $\gtrsim 400$ & 415 &       2 &     2 &      39 & $<1$
                 & \nodata & $<1$ \nl
\OII \lamOII     & \nodata       & 118 &      10 &     9 &      52 &    6
                 &      12 &   12 \nl
\Hb              & \nodata       &  48 & \nodata &     1 &      13 &  1.5
                 & \nodata &    2 \nl
\Ha              & \nodata       & 147 & \nodata &     4 & \nodata &    5
                 & \nodata &    6 \nl
\MgII \lamMgII   & \nodata       &  94 & \nodata &     3 & \nodata &    3
                 & \nodata &    5 \nl
\NeIII \lamNeIII & \nodata       &  42 & \nodata &  $<1$ & \nodata & $<1$
                 & \nodata & $<1$ \nl
\NII \lamNII     & \nodata       &  32 & \nodata &     4 & \nodata &    4
                 & \nodata &    6 \nl
\Hg              & \nodata       &  27 & \nodata &  $<1$ & \nodata & $<1$
                 & \nodata &    1 \nl
\cutinhead{Parameters of ionization models}
$n_H$ (cm$^{-3}$) & & 0.3 & & 0.83 & & 1 & & 7.5 \nl
$M_{gas}$ ($M_{\sun}$) & & $<8\times 10^7$ & & $<10^{10}$ & & \nodata & &
   $<4\times 10^{10}$
\enddata
\tablerefs{[1] \cite{SM:87}; [2] \cite{B:96};
           [3] \cite{W:94};  [4] \cite{MS:85}.}
\end{deluxetable}


\end{document}